\documentclass{ws-ijmpe}

\begin{document}

\markboth{M.A. Mardyban, V.O. Nesterenko,  P.-G. Reinhard, A. Repko and J. Kvasil}
{Moments of inertia in light deformed nuclei: pairing and mean-field impacts}


\title{Moments of inertia in light deformed nuclei: pairing and mean-field impacts}
\author{ V.O. Nesterenko $^{1,2}$, M.A. Mardyban $^{1,2}$, P.-G. Reinhard
$^{3}$, A. Repko $^{4}$ and J. Kvasil $^{5}$}
\address{$^1$
  Laboratory of Theoretical Physics, Joint Institute for Nuclear Research.
  141980, Dubna, Moscow region, Russia}
\address{$^2$
   Dubna State University. 141982, Dubna, Moscow region, Russia}
\address{$^3$
     Institute for Theoretical Physics II, University of Erlangen, D-91058, Erlangen, Germany}
\address{$^4$
     Institute of Physics, Slovak Academy of Sciences, 84511 Bratislava, Slovakia}
 \address{$^5$
 Institute of Particle and Nuclear Physics, Charles
     University, CZ-18000, Praha 8, Czech Republic}

\maketitle

\begin{history}
\received{(received date)} \revised{(revised date)}
\end{history}

\begin{abstract}
   The dependence of the moment of inertia $\cal J$ on the pairing and axial quadrupole deformation
   $\beta$ in $^{24}$Mg and $^{20}$Ne was investigated.
   The study is based on quadrupole-constrained calculations with
  three cranking approaches for $\cal J$ (Inglis-Belyaev, Thouless-Valatin, adiabatic
  time-dependent Hartree-Fock) and a representative set of Skyrme forces (SVbas, SkM*, SLy6).
  At variance with macroscopic collective models, the calculations predict the specific
  regime $d{\cal J}/d\beta<0$ at $\beta \ge 0.5$ ($^{24}$Mg) and $\beta \ge 0.6$
  ($^{20}$Ne), where the pairing breaks down. This regime is explained by two effects:
  full break up of the pairing and specific
evolution of a {\it single} dominant particle-hole (1ph) configuration with
$\beta$.  The analysis of experimental data for the ground-state rotational bands in $^{24}$Mg and
$^{20}$Ne shows that such regime is possible at low spins.
\end{abstract}

\maketitle

\newcommand{\vek}[1]{\mathbf{#1}}
\newcommand{\me}{\mathrm{e}}
\newcommand{\mi}{\mathrm{i}}
\newcommand{\radi}{(\vek{r})}
\newcommand{\radii}{\vek{r}}
\newcommand{\der}[2]{\mathrm{d}^{#2} #1 \,}
\newcommand{\bra}[1]{\langle #1 |}
\newcommand{\ket}[1]{| #1 \rangle}
\newcommand{\squary}[1]{\left[ #1\right]}
\newcommand{\wavy}[1]{\left( #1\right)}

\section{Introduction}

  The moment of inertia is one of the most important characteristics of deformed atomic nuclei
\cite{BM2,Ring,Sol,Sit}. As was shown by extensive theoretical and experimental studies
for medium and heavy nuclei, moments of inertia $\cal J$ generally grow with nuclear
quadrupole deformation $\beta$ (regime $d{\cal J}/d\beta >0$) and decrease with a development of pairing.
The first feature is demonstrated by familiar macroscopic rigid-body (RB) and
hydrodynamical (HD) estimations for $\cal J$ \cite{BM2,Ring,Sol,Sit}. The
regime $d{\cal J}/d\beta >0$ is mainly relevant for the collective nuclear flow
realized in medium and heavy nuclei.
Light nuclei with their extreme deformations and strong shell effects
suggest additional interesting opportunities for investigation of dependence of $\cal J$ on
the deformation, shell structure and pairing.

In this study, we show that the usual trend that $\cal J$ grows
with $\beta$ can be reverted in light deformed nuclei $^{24}$Mg and $^{20}$Ne,
i.e., we can get in these nuclei the regime  $d{\cal J}/d\beta <0$. This regime
can be governed  by a {\it single} 1ph configuration and realized at zero pairing.

There are many cases when interplay of a collective motion
and 1ph or two-quasiparticle (2qp) excitations essentially
affects moments of inertia. These cases include various
shell-corrections  \cite{RMP72}, backbending \cite{Ring,bb},
onset of nuclear triaxiality \cite{BM2,Mar,Naz,Frauendorf,Kva06}, etc.
The effects were mainly explored  within various versions
of the unified collective rotor model of Bohr and Mottelson \cite{BM2,Ring}
and cranking model (CM) originally proposed by D.R. Inglis \cite{In54,In56} (see
\cite{Ring,Frauendorf} for extensive CM reviews). In some CM studies, one can
find cases with  $d{\cal J}/d\beta < 0$, see early \cite{PF75} and more
recent \cite{Kva06} examples. However, these cases mainly concern medium/heavy
nuclei and high-spin regimes. At the same time, it would be interesting to find
mean-field effect in $\cal J$ beyond the cases mentioned above, e.g. for low spins,
without band crossing and even without pairing impact. Light deformed
nuclei look promising for this aim.

Various properties of light nuclei $^{24}$Mg and $^{20}$Ne,
including their moments of inertia, were  explored already for many decades,
see e.g. review~\cite{BM2} for the early work.
In particular, the thorough analysis (CM with Nilsson-Strutinsky formalism) of the spectra,
axial/triaxial deformation paths and moments of inertia
in rotating sd-shell nuclei was performed by Lund group in 1980s \cite{RA81}.
The effect of pairing in rotating $^{24}$Mg was investigated \cite{MosPLB,MosNPA}.
Nevertheless, despite an impressive previous effort, low-energy spectroscopy
of $^{24}$Mg and $^{20}$Ne remains to be a hot topic. For example, during last years,
the impact of triaxiality and shape coexistence was revisited within various  methods:
Skyrme quasiparticle random-phase-approximation (QRPA) \cite{Yoshida_PRC08,Losa_PRC10},
the constrained Hartree-Fock-Bogoliubov + local QRPA (CHFB+LQRPA) method\cite{Hinohara_PRC11},
generator coordinate method with angular-momentum-projected  triaxial relativistic
mean-field wave functions (3DAMP+GCM) \cite{Yao_PRC11},
GCM with full triaxial angular momentum and particle number projection using Skyrme \cite{Bender_PRC08}
and  Gogny \cite{Rod_PRC10} forces,
triaxial CM \cite{Gul22},
Antisymmetrized Molecular Dynamics (AMD) \cite{20Ne_PRC04,24Mg_Kim,Chi21}, etc. It was shown that
rotational bands built on even-parity excited states in $^{24}$Mg and $^{20}$Ne can exhibit some
triaxility, and this effect is most strong in the soft nucleus  $^{20}$Ne. Further, the
influence of clustering was inspected \cite{20Ne_PRC04,24Mg_Kim,Chi21,BI20}. The low-energy
 spectrum of $^{24}$Mg was recently measured in a nuclear-resonance-fluorescence (NRF) experiment
 \cite{exp22}. To our knowledge, the regime $d{\cal J}/d\beta <0$ was actually found only
in one study~\cite{Hinohara_PRC11} (as a part of a local ${\cal J}(\beta)$-maximum caused by
a pairing collapse) but a possible mean-field origin of this feature was not explored.

In this paper, we consider deformed nuclei $^{24}$Mg and $^{20}$Ne as promising candidates
where the regime $d{\cal J}/d\beta <0$ can be apparently realized.
The fully self-consistent approaches with Skyrme forces are employed to
analyze a mean-field origin of  $d{\cal J}/d\beta <0$ effect in terms of single-particle spectra.
Our results suggest a simple microscopic interpretation of the effect, which can be useful for
understanding and treatment of more involved explorations.

The paper is organized as follows. In Sec. II, various methods for calculation of
moments of inertia are sketched. In Sec. III, the calculation details are outlined. In Sec. IV,
the main results are presented and analyzed, the experimental status
is discussed. In Sec. V, the conclusions are drawn. In Appendix A, the case of a weak pairing
is illustrated. In Appendix B, properties of dominant 2qp excitations are exhibited.

\section{Models for moments of inertia}

The moment of inertia $\cal J$ is usually
defined through the expression
for the rotational energy \cite{BM2,Ring,Sol,Sit}
\begin{equation}
E_{I}=\frac{\hbar^2}{2\cal J}I(I+1),
\label{E_I}
\end{equation}
where $I$ is the angular momentum of the rotational state.

The ${\cal J}$ can be modeled in several ways.
Familiar {\it macroscopic} approaches~\cite{BM2,Ring,Sol,Sit} suggest
the rigid-body (RB)
 \begin{equation} 
{\cal J}_{\rm{RB}}=\frac{2}{5} M R^2 \left(1+\frac{1}{2}
\sqrt{\frac{5}{4\pi}}\beta+\frac{25}{32\pi}\beta^2\right)
\label{RB}
\end{equation}
and hydrodynamical (HD)
 \begin{equation} 
{\cal J}_{\rm{HD}}=\frac{9}{4\pi} M R^2
\frac{\beta^2 (1+\frac{1}{4}\sqrt{\frac{5}{4\pi}}\beta)^2}
{2+\sqrt{\frac{5}{4\pi}}\beta + \frac{25}{16\pi}\beta^2}
\;,
\label{HD}
\end{equation}
expressions. Here, $M$ and $R$ are the nuclear mass and radius, $\beta$ is the dimensionless
axial quadrupole deformation. Pairing  in these formulas is absent.
At a low deformation $\beta <$ 0.4, we get the relation
\begin{equation} 
{\cal J}_{\rm{HD}}={\cal J}_{\rm{RB}}\frac{45}{16\pi}\beta^2 ,
\label{HD-RB}
\end{equation}
where ${\cal J}_{\rm{RB}} > {\cal J}_{\rm{HD}}$.
For rare-earth and actinide nuclei, experimental values of $\cal J$
usually lie between the RB and HD estimations.
Both RB and HD expressions predict a  growth of $\cal J$ with deformation.
Macroscopic RB and HD models describe moment of inertia for a collective rotation.
If we are interested in mean field effects for ${\cal J}$, we should consider
microscopic  models.

A {\it microscopic} expression for $\cal J$ is given by the
Inglis cranking formula \cite{Ring,In54,In56}
\begin{equation}
\label{Inglis}
{\cal J}_{\rm Ing}=2\sum_{i}\frac{|\langle i|I_x|0\rangle|^2}{E_i} ,
\end{equation}
where $I_x$
is x-component of the operator of the total angular moment $I$, $|0\rangle$
is the ground state (g.s.), and $|i\rangle$ is an excited state with energy
$E_i$ and quantum numbers $K^{\pi}=1^+$ ($K$ is  projection of $I$
onto the nuclear symmetry axis, $\pi$ is the parity.)

Expression (\ref{Inglis}) can be used at different levels of complexity. In the
simplest mean-field case, we deal with mere particle-hole (ph) excitations. Then
$|i\rangle = |ph\rangle$ and $E_i=e_p-e_h$, where $e_p$ and  $e_h$ are energies of the
particle and hole single-particle levels. This yields the original
Inglis cranking formula  \cite{In54,In56}.
If the pairing is included, we obtain the Inglis-Belyaev (IB) formula  \cite{Be59,Be61}
\begin{equation}
{\cal J}_{\rm IB}=2\sum_{q,q'>0} \frac{|\langle qq'|I_x|\tilde{0}\rangle|^2}
{\epsilon_q+\epsilon_{q'}} ,
\label{IB}
\end{equation}
where $|\tilde{0}\rangle$ is the quasiparticle vacuum.
Here we deal with 2qp states
$|i\rangle = |qq'\rangle$ with $q,q' >0$ and excitation energy $E_i=\epsilon_q+\epsilon_{q'}$.
The 2qp matrix element
 $\langle qq'|I_x|\tilde{0}\rangle$ includes the pairing weight factor $u_{qq'}=u_q v_{q'}-u_{q'} v_q$
with Bogoliubov coefficients $u_q$ and $v_q$.

  A further improvement is  QRPA which takes  also the residual interaction into account.
Then we get the Thouless-Valatin (TV) expression \cite{TV62}
\begin{equation}
\label{TV}
{\cal J}_{\rm TV}=2\sum_{\nu} \frac{|\langle\nu|I_x|\tilde{\tilde{0}}\rangle|^2}{E_{\nu}} ,
\end{equation}
where $|\tilde{\tilde{0}}\rangle$ is the QRPA vacuum, $|\nu\rangle$ is
the excited QRPA state with the energy $E_{\nu}$.

It is easy to see that Eqs. (\ref{Inglis})-(\ref{TV})
allow various trends of ${\cal J}$ with deformation. For example,
if deformation $\beta$ only slightly affects squared matrix elements
$|\langle i|I_x|0\rangle|^2$ but leads to a large change of the energy
$E_i$, then one can get any sign for $d{\cal J}/d\beta$. We will
show below that a negative sign can occur in $^{20}$Ne and
$^{24}$Mg.

Note that matrix QRPA can be unstable and not sufficiently accurate at
deformations far from the equilibrium values. So Eq. (\ref{TV}) which uses
the QRPA output is not always a reliable tool to estimate ${\cal J}$ at all
points of the deformation path. To circumvent this limitation, we also use in
our analysis the linear adiabatic time-dependent Hartree-Fock (ATDHF) approach
\cite{Ring,ATDHF_GR78,ATDHF_RGG84}  where ${\cal J}$ is calculated  directly
through the 2qp operator $\Theta_x(\beta)$ of the linear response to the
perturbation $I_x(\beta)$:
\begin{eqnarray}
\label{ATDHF}
 [\Theta_x(\beta), H(\beta)]&=&
 -i\hbar \frac{I_x(\beta)}{{\cal J}_{\rm ATDHF}(\beta)},
 \\
{\cal J}_{\rm ATDHF} (\beta) &=&
\frac{\hbar^2}{\langle \Phi_{\beta} |[[\Theta_x, H],\Theta_x]| \Phi_{\beta}\rangle} ,
\end{eqnarray}
where $H(\beta)$ and  $\Phi_{\beta}$ are Hamiltonian  and ground state of the system
at a given $\beta$.

Formally, the ATDHF and TV definitions should result in the same ${\cal J}$.
However, in our study, these two models use different numerical
realizations. In particular, ATDHF evaluates the linear response iteratively in
a two-dimensional coordinate grid~\cite{ATDHF_RGG84}, which is more
robust than the matrix QRPA technique exploiting a finite expansion basis.
In our analysis, we mainly use ATDHF.
The application of TV is limited to a few deformation points where
QRPA has stable solutions just to show that TV and ATDHF give very similar results.

\section{Calculation details}

The calculations are performed with Skyrme parametrizations SVbas \cite{SVbas},
SkM* \cite{SkMs} and SLy6 \cite{SLy6}, which have different isoscalar effective masses
 $m^*/m$ (0.90, 0.79 and 0.69, respectively). It is known that the smaller $m^*/m$, the more
 stretched the single-particle spectrum. Thus $\cal J$
can depend on $m^*/m$. Besides, these parametrizations use different prescriptions
for pairing \cite{pair}: surface (density-dependent) pairing for
SVbas and volume (density-independent) pairing for SkM* and SLy6.
Pairing is treated within Bardeen-Cooper-Schrieffer (BCS) scheme~\cite{pair,Rep_EPJA17}.
The center-of-mass corrections are computed following
the prescriptions for the ground \cite{BHR03} and excited \cite{Kva_sebrpaEPJA19} states.

\begin{table} 
\caption{
Calculated equilibrium deformations $\beta_{\mathrm{eq}}$,
 proton $\Delta_p$ and neutron $\Delta_n$
pairing gaps, and ATDHF moments of inertia $\cal J$ in
$^{24}$Mg and $^{20}$Ne.}
\begin{tabular}{|c|c|c|c|c|c|}
\hline
Nucleus     &                      & exper \cite{nndc} & SVbas & SkM* & SLy6  \\
\hline
$^{24}$Mg & $\beta_{\mathrm{eq}}$ & 0.61 & 0.52 & 0.49 & 0.54 \\
          &$\Delta_p$ [MeV]       &      &  0.59 & 0.01 & 0.01 \\
          &$\Delta_n$ [MeV]       &      &  0.02 & 0.01 & 0.01 \\
          & ${\cal J}$ $[\hbar^2$/MeV]     & 2.2  &  3.6 & 4.1 & 3.5 \\
\hline
$^{20}$Ne & $\beta_{\mathrm{eq}}$ & 0.72 &  0.31 & 0.37 & 0.56 \\
          &$\Delta_p$ [MeV]       &      &   1.87 & 1.15 & 0.02 \\
          &$\Delta_n$ [MeV]       &       &  1.63 & 1.07 & 0.16\\
          & ${\cal J}$ $[\hbar^2$/MeV]     & 1.8 &  0.7 & 2.0 & 5.8 \\
\hline
\end{tabular}
\label{perf}
\end{table}

The equilibrium deformations $\beta_{\rm eq}$ are obtained by minimization of
the total nuclear energy. Following  Table~\ref{perf}, the calculated $\beta_{\rm eq}$
underestimate the experimental values. This is explained by shallow
potential energy surfaces (PES) in these nuclei, especially in very soft $^{20}$Ne.
Underestimation of $\beta_{\rm eq}$ in light nuclei also takes place in other
calculation schemes, e.g. CHFB+LQRPA model with SkM* gives $\beta_{\rm eq}$=0.41
in $^{24}$Mg \cite{Hinohara_PRC11}.
Similar values for $\beta_{\rm eq}$ in $^{24}$Mg  are also obtained in some
density functional theories \cite{Yoshida_PRC08,Losa_PRC10,Yao_PRC11,Bender_PRC08}.
For $^{20}$Ne, the value $\beta_{\rm eq}\sim$0.41 is obtained in AMD scheme~\cite{20Ne_PRC04}.
Note that present calculations of PES do not take
into account the rotational and vibrational zero-point energy (ZPE)
corrections \cite{ZPA74}. Following our estimations,
these corrections could strengthen the deformed minima and significantly
improve agreement of $\beta_{\rm eq}$ with $\beta_{\rm exp}$.
However, a thorough analysis of $\beta_{\rm eq}$ is beyond the present study
where we are  mainly interested in the constrained calculations over a wide
$\beta$-range.
Note also that $\beta_{\rm exp}$ are obtained from the
$B(E2)$ values for ground-state (g.s.) rotational bands.  In soft nuclei, $B(E2)$
include additional large dynamical correlations \cite{ATDHF_RGG84} leading to
overestimation of extracted $\beta_{\rm exp}$.

\begin{figure} 
 \centering
 \includegraphics[width=0.40\textwidth]{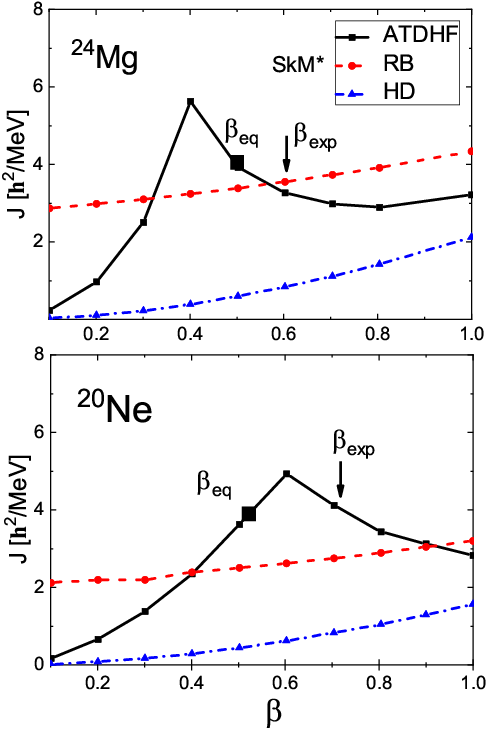}
\caption{Moments of inertia  in $^{24}$Mg and $^{20}$Ne, calculated
within RB, HD and SkM* ATDFT models. The equilibrium and experimental deformations
are  marked by the filled rectangles and arrows, respectively.}
\label{fig:atdhf_rb_hd}
\end{figure}

In the IB and ATDHF models, we calculate $\cal J$ using
the code SKYAX \cite{Rein}. A two-dimensional (2D) grid in cylindrical
coordinates with grid step 0.7 fm and calculation box up to 3 nuclear
radii is employed. All proton and neutron single-particle (s-p) levels from the bottom of the
potential well up to +40 MeV  are included. For example, SkM* calculations
at equilibrium deformations in $^{24}$Mg employ 1050/1050 proton/neutron s-p levels.
All 2qp $K^{\pi}=1^+$ states until 60 MeV are included, e.g. in $^{24}$Mg we use
1770/1770 proton/neutron 2qp configurations. To make sure that the size of the
expansion  basis suffices, we check the full (lm=20,21,22) quadrupole energy-weighted
sum rule $\textrm{EWSR(E2)}=\hbar^2 e^2/(8\pi m) 50 Z \langle r^2 \rangle_Z$
(where $m$ is the nucleon mass, $Z$ is the nuclear charge and  $\langle r^2 \rangle_Z$ is
$r^2$-averaged  proton density). The EWSR is exhausted by 92-100$\%$.

 In TV cranking, we employ fully self-consistent 2D matrix QRPA method
\cite{Repko_arxiv} using s-p spectra and pairing values from SKYAX code.
The spurious admixtures in QRPA spectra are additionally extracted
\cite{Kva_sebrpaEPJA19}.

The experimental $\cal J$-values are obtained from  Eq.~(\ref{E_I})
using  energies  of $2^+$ states in the  g.s. rotational band \cite{nndc}.

\section{Results and discussion}
 \subsection{Trends with deformation}

 Fig.~\ref{fig:atdhf_rb_hd} exhibits evolution ${\cal J}(\beta)$
 in $^{24}$Mg and $^{20}$Ne, calculated  within RB, HD and SkM* ATDHF models.
 We see that, in both nuclei,  RB and HD predict a gradual increase of
 $\cal J$ with $\beta$. Instead, ATDHF produces an maximum in ${\cal J}(\beta)$
 at $\beta\sim$ 0.4 in $^{24}$Mg and  $\beta\sim$ 0.6 in $^{20}$Ne with a subsequent
 dramatic decrease of $\cal J$ at larger $\beta$ (regime $d{\cal J}/d\beta <0$).
 To understand this counterintuitive behavior, we present below a detailed microscopic analysis.

\begin{figure*} 
\centering
\includegraphics[width=0.8\textwidth]{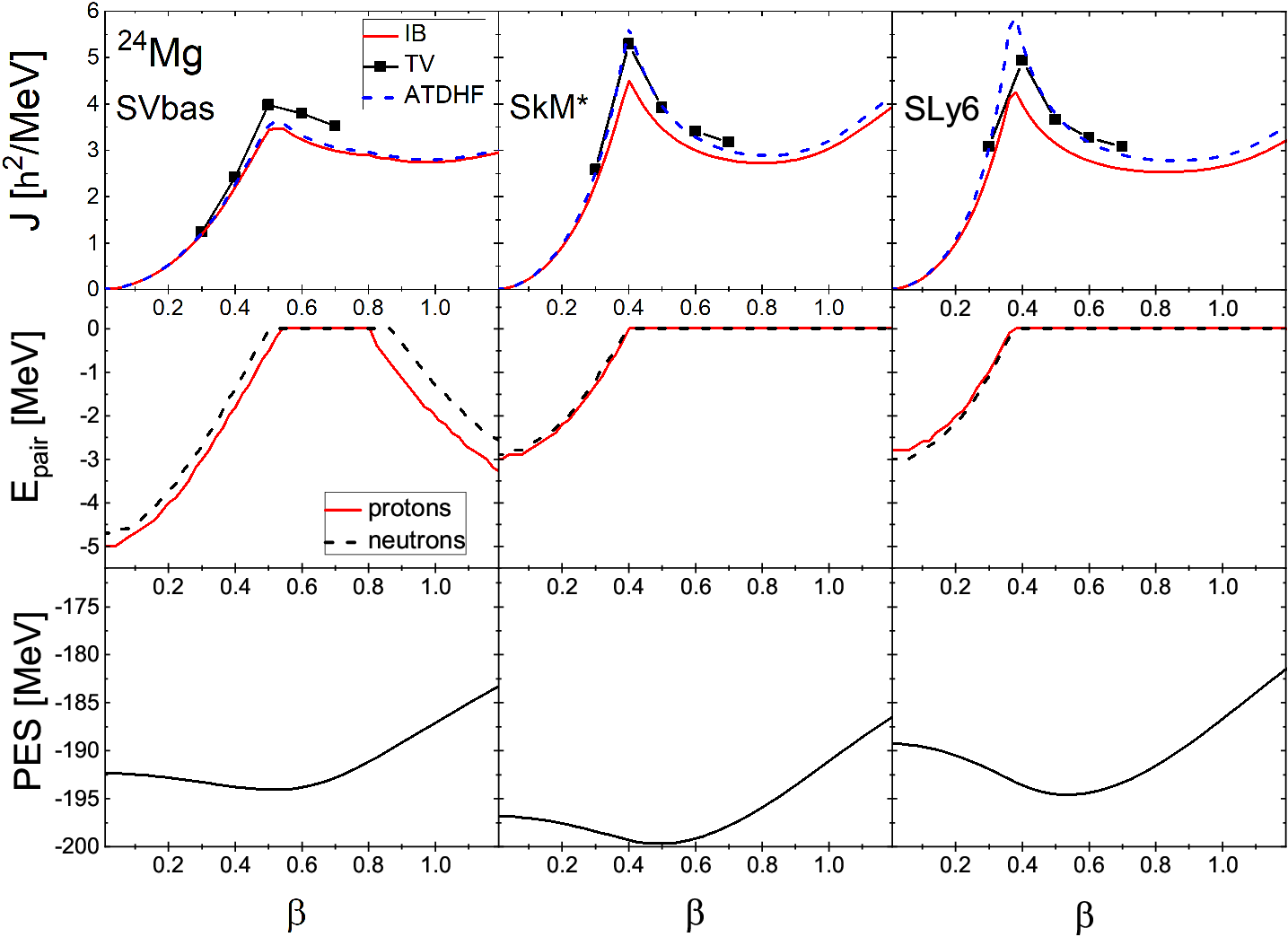}
\caption{Moments of inertia $\cal J$ (upper panels),  neutron and proton pairing energies
$E_{\rm pair}$ (middle panels) and potential energy surfaces PES (bottom panels) for $^{24}$Mg, calculated
with the forces SVbas, SkM* and SLy6. $\cal J$-values are obtained within IB, TV and
ATDHF models.}
\label{fig2:Mg}
\end{figure*}

Fig.~\ref{fig2:Mg} shows ${\cal J}(\beta)$ in $^{24}$Mg, calculated within
the IB, TV, and ATDHF microscopic models for different Skyrme forces.
TV results are given only for a few deformation points of our main interest.
Besides, we demonstrate neutron and proton pairing energies $E_{\rm pair}$ (defined in
Refs. \cite{pair,Rep_EPJA17}) and PES. The values of $E_{\rm pair}$  and PES are the same
for IB, TV and ATDHF.

The upper panels of Fig.~\ref{fig2:Mg} show that all three microscopic
models give similar results for $\cal J$. This means that QRPA correlations provided
by TV and ATDHF are not important. At low deformations, we see  a gradual growth of
$\cal J$, which is mainly caused by  decrease of the pairing.
The pairing impact overrides another factor - a change of 2qp energies
$\epsilon_q+\epsilon_{q'}$ in the denominator of ${\cal J}_{\rm RB}$.
At $\beta >$ 0.5 (SVbas) and $\beta >$ 0.4 (SkM*, SLy6), the pairing
fully disappears (in accordance with calculations \cite{Losa_PRC10,Hinohara_PRC11,Yao_PRC11}
where a collapse of pairing is found at similar $\beta$-values), which makes the energy factor
decisive. At these deformations, the energy denominator increases with $\beta$ (see discussion
below) and thus we get a gradual decrease of $\cal J$. The effect becomes fully of the mean field origin.
The regime $d{\cal J}/d\beta <0$ is more pronounced for SkM* and SLy6 (parametrization with a small $m^*/m$),
where $\cal J$ falls down almost twice. For even larger deformations, the trend turns back to
a usual growth of $\cal J$ with $\beta$.

\begin{figure*}
\centering
\includegraphics[width=0.8\textwidth]{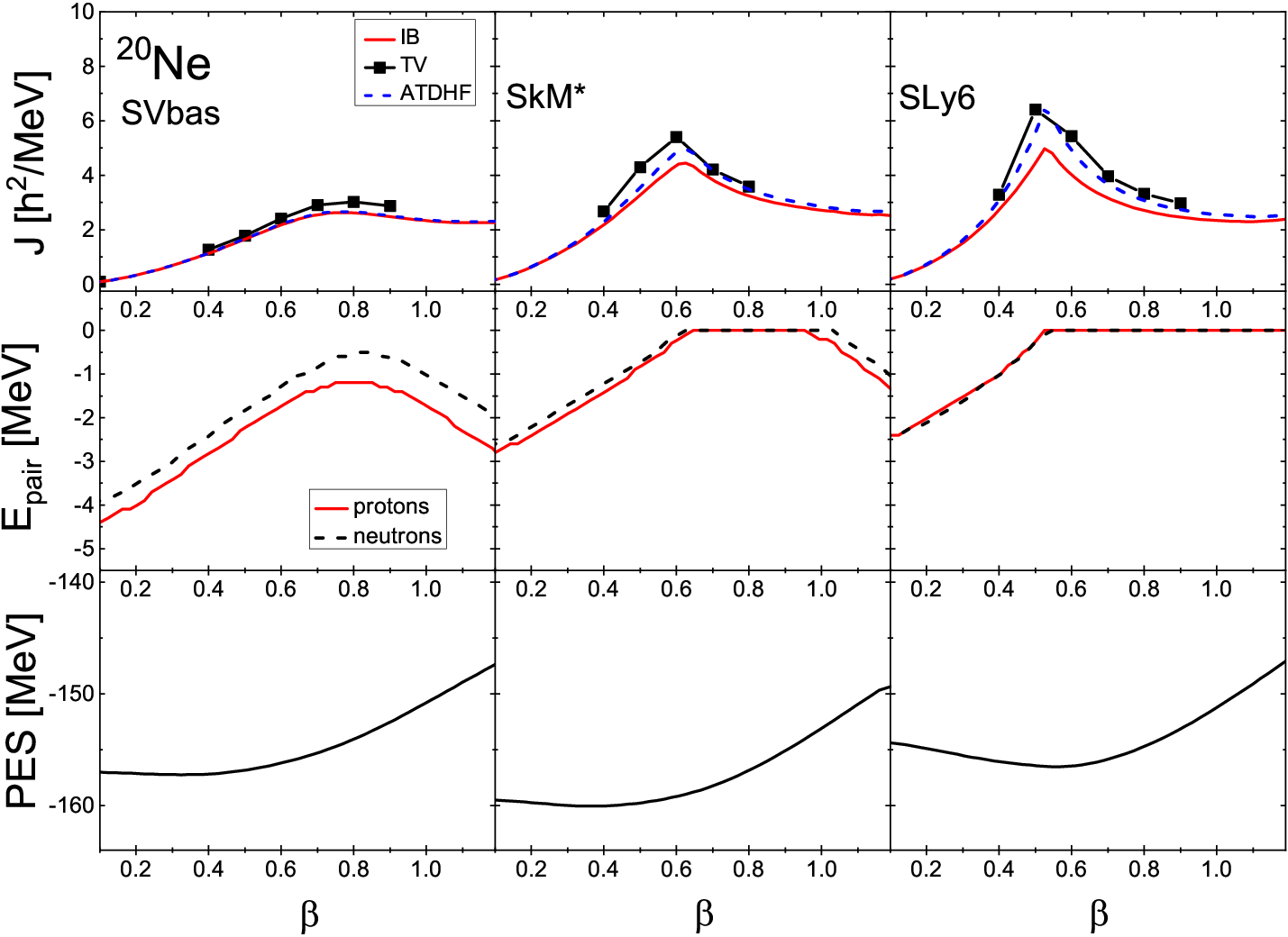}
\caption{The same as in Fig.~\ref{fig2:Mg} but for $^{20}$Ne.}
\label{fig3:Ne}
\end{figure*}

Fig.\ref{fig3:Ne} shows the same trends for $^{20}$Ne. In both $^{24}$Mg and $^{20}$Ne, the effect
is less pronounced for SVbas, i.e.  for the force with a more developed pairing.
The pairing smooths the ${\cal J}(\beta)$-maximum. A particle-number projection
(as an improvement of BCS method) could lead in our cases to some
onset of pairing \cite{Bender_PRC08,Rod_PRC10,Erl08a}.

As mentioned above, the maximum in ${\cal J}(\beta)$ at $\beta$=0.4-0.5 in $^{24}$Mg was also found
in CHFB+LQRPA calculations taking into account the triaxiality and rotation~\cite{Hinohara_PRC11}.
Altogether, ${\cal J}(\beta)$-maximum was obtained in prolate $^{24}$Mg, oblate  $^{28}$Si and
shape-mixed $^{26}$Mg and $^{24}$Ne~\cite{Hinohara_PRC11}.  So this phenomenon could be rather common
for light deformed nuclei. The study~\cite{Hinohara_PRC11} shows that i) ${\cal J}(\beta)$-maximum
in $^{24}$Mg persists in the g.s. band while increasing the angular momentum $I$ and ii) this effect
is basically caused by the collapse of proton and neutron pairing at $\beta$=0.4-0.5.
The latter  is in accordance with our calculations where just a collapse of pairing
makes the energy impact decisive and leads to regime $d{\cal J}/d\beta <0$. At $\beta >$0.5, this
regime occurs at zero pairing and, as shown below, is governed by evolution of a single 1ph
configuration  with $\beta$.

\subsection{Microscopic analysis of the results}

Since IB, TV and ATDHF  give similar results, we use for our analysis only
IB model which, being the most simple, embraces nevertheless
the most important mean field and pairing impacts. For the sake of brevity, we
limit our inspection by SkM* results for $^{24}$Mg.

\begin{figure} 
\centering
\includegraphics[width=0.5\textwidth]{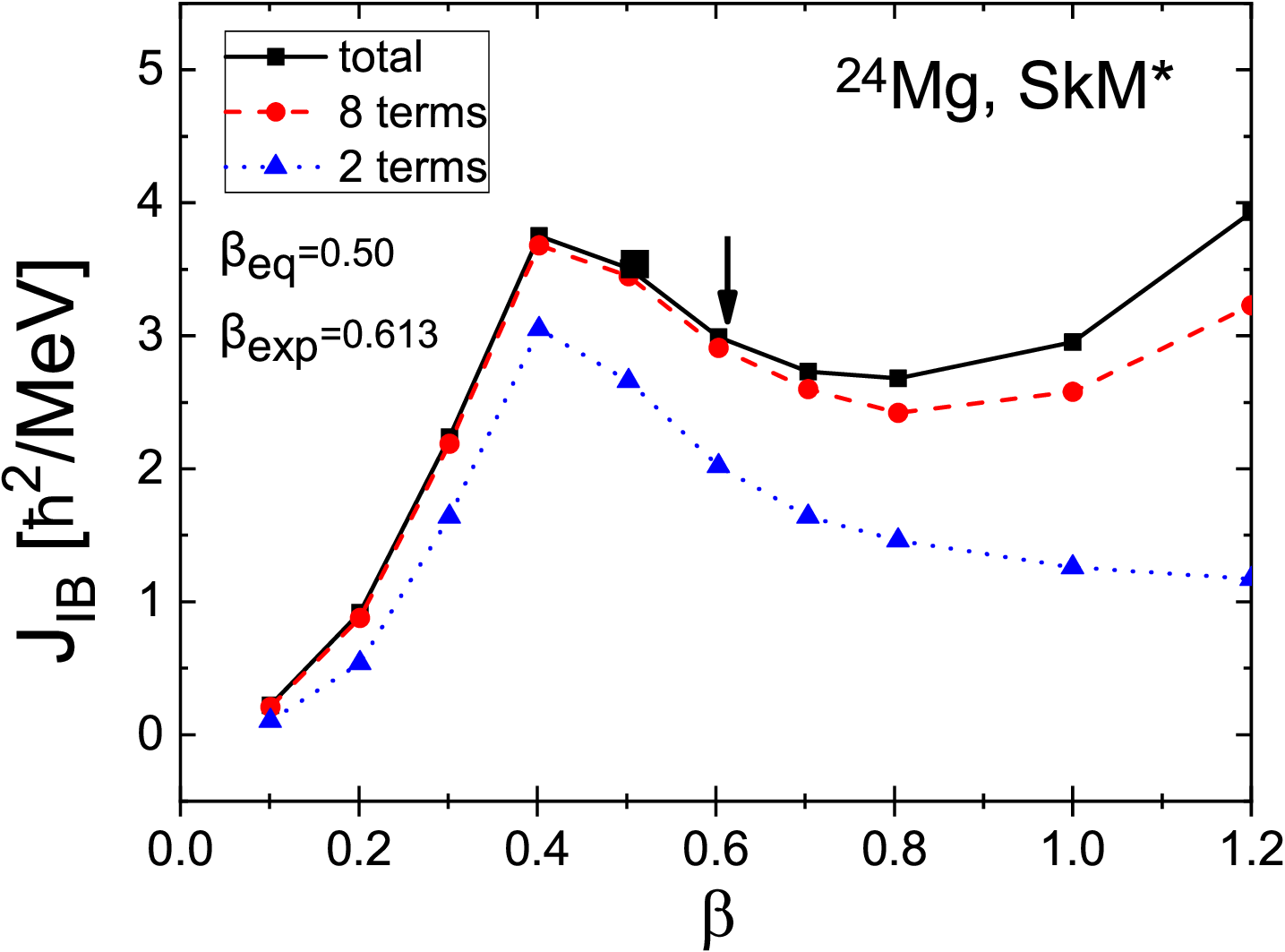}
\caption{$\cal J_{\rm IB}$ in $^{24}$Mg, calculated with the full 2qp set and
limited numbers (8 and 2) of 2qp terms.  The equilibrium and experimental deformations
are  marked by the filled rectangle and arrow, respectively.}
\label{Mg_Jqp}
\end{figure}

Fig.\ref{Mg_Jqp} shows $\cal J_{\rm IB}(\beta)$ in $^{24}$Mg for
the full and limited 2qp configuration spaces. In the latter case,
two and  eight 2qp pairs with maximal contributions  to  $\cal J_{\rm IB}$
(Eq. (\ref{IB})) are taken into account. It is seen that eight 2qp configurations
(4 proton and similar 4 neutron pairs) reproduce the behavior
of the full $\cal J_{\rm IB}$ rather well. Even two configurations
(proton and similar neutron) show the main $d{\cal J}/d\beta < 0$ effect.

Table~\ref{tabMg} (Appendix B) collects some properties (energies,
squared $I_x$-matrix elements, pairing factors) of
 2qp pairs dominating in Eq. (\ref{IB}) for $\cal J_{\rm IB}$.
The states are characterized by asymptotic Nilsson quantum numbers
$[N,n_z,\Lambda]$ \cite{Nilsson65} with arrows indicating direction of
the spin.  The table shows that two configurations, proton and neutron
$[211\uparrow, 202\uparrow]$, give the dominant contribution to $\cal
J$ at all the considered deformations.

 \begin{figure*} 
\centering
\includegraphics[width=0.75\textwidth]{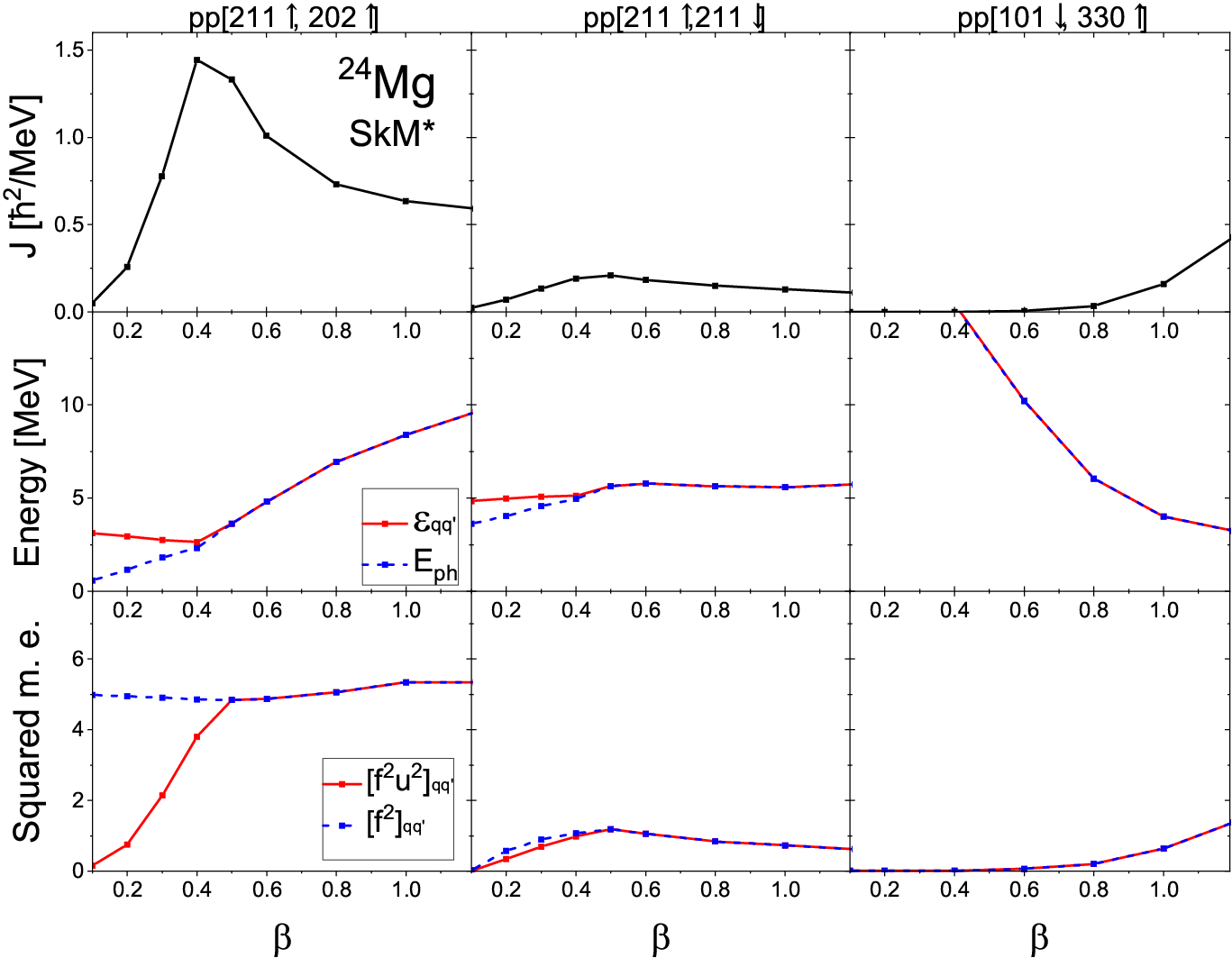}
\caption{Contributions ${\cal J}_{qq'}$ to the
moment of inertia ${\cal J}_{\rm IB}$, 2qp energies $\epsilon_{qq'}$, 1ph energies
$E_{\rm ph}$, squared single-particle matrix elements  $f^2_{qq'}$ and products
$(f_{qq'} u_{qq'})^2$ for pairs
pp$[211\uparrow, 202\uparrow]$, pp$[211\uparrow, 211\downarrow]$
and pp$[101\downarrow, 330\uparrow]$ in $^{24}$Mg.}
\label{Mg_2qp_frac}
\end{figure*}

Left panels of Fig.~\ref{Mg_2qp_frac} illustrate $\beta$-dependence of some characteristics
of the proton excitation pp$[211\uparrow, 202\uparrow]$. We  show the contribution
\begin{equation}
  {\cal J}_{qq'}= 2 \frac{|f_{qq'} u_{qq'}|^2}{\epsilon_{qq'}}
  \label{Jqq}
\end{equation}
to ${\cal J}_{\rm IB}$  and all the values entering this
expression: 2qp energy $\epsilon_{qq'}=\epsilon_{q}+\epsilon_{q'}$ or
particle-hole energy $E_{qq'}$ (for zero pairing),
single-particle matrix element $f_{qq'}=\langle q|I_x|q' \rangle$
and pairing weight $u_{qq'}$. Only the proton pair is
considered since, in $N=Z$ nuclei like $^{24}$Mg, features of neutron
and proton excitations are similar (see Table~\ref{tabMg}).
Fig.~\ref{Mg_2qp_frac} shows that pairing is important for
$\beta <$ 0.5. In this deformation range, $\epsilon_{qq'}> E_{ph}$,
$(f_{qq'})^2 > (f_{qq'}u_{qq'})^2$ and ${\cal J}_{qq'}$ grows with $\beta$.
At higher deformations, pairing vanishes, the energy $E_{ph}$ gradually increases
and the squared matrix element $(f_{qq'})^2$ is almost constant. Such
behavior of the energy and matrix element results in the decrease
of  ${\cal J}_{qq'}$ with $\beta$. Since proton and neutron configurations
$[211\uparrow, 202\uparrow]$ strongly dominate in ${\cal J}_{\rm IB}$,
we finally get the anomalous decrease of ${\cal J}_{\rm IB}$ with
$\beta$. This trend stems exclusively from shell effects.


\begin{figure*} 
\centering
\includegraphics[width=0.65\textwidth]{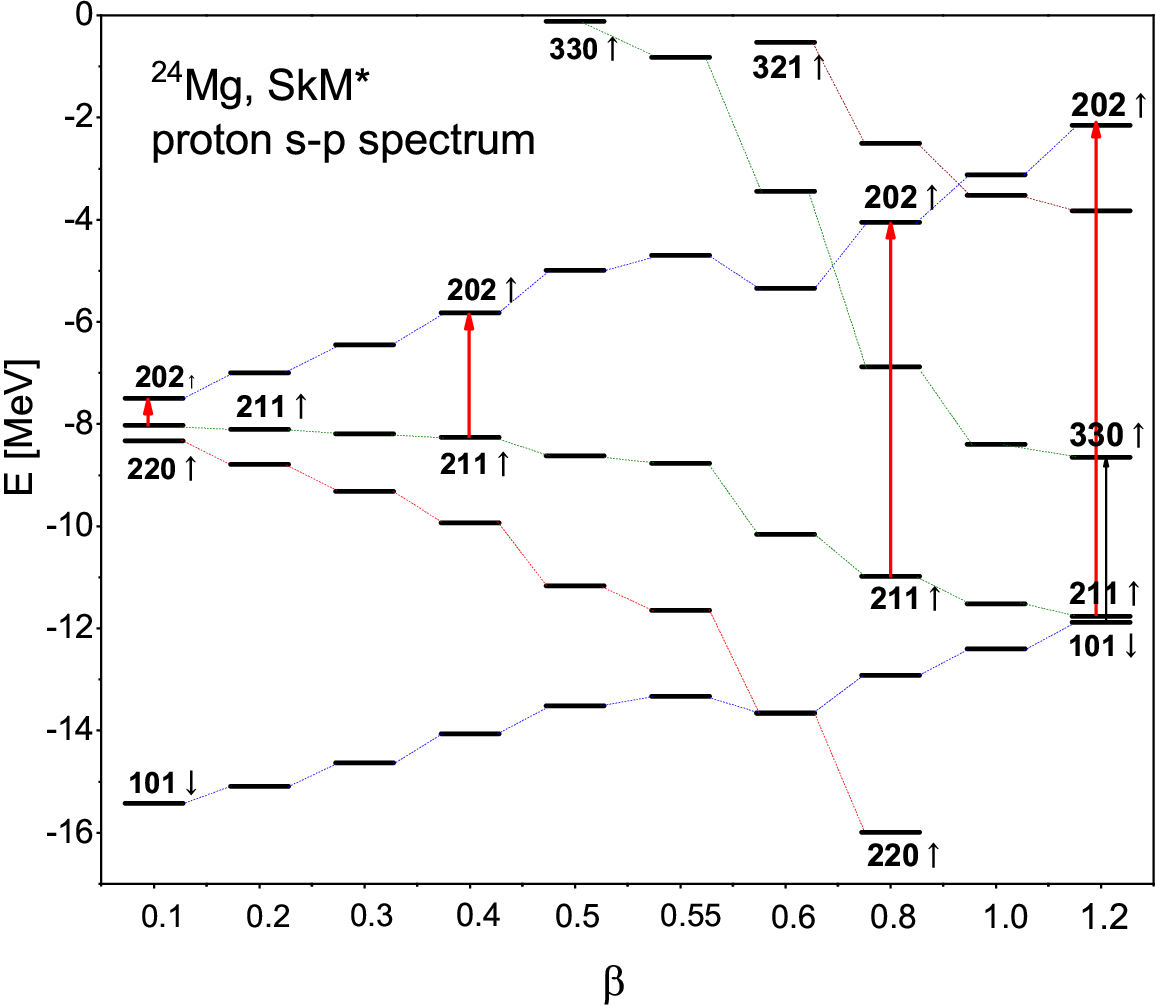}
\caption{Proton single-particle SkM* spectrum in $^{24}$Mg. The
transition $[211\uparrow \to 202\uparrow]$ is shown by red arrows.}
\label{24Mg_sps}
\end{figure*}


The middle and right panels of Fig.~\ref{Mg_2qp_frac} show the similar
values for pairs pp[211$\uparrow$, 211$\downarrow$] and
pp[101$\downarrow$, 330$\uparrow$]. As compared with left panels, they
demonstrate different evolution  of the energies and matrix elements.
At 0.5 $< \beta <$ 0.8, their matrix elements
are small, which results in low contributions to ${\cal J}_{qq'}$.

The behavior of the energy  $\epsilon_{qq'}$ for the dominant pair
$[211\uparrow, 202\uparrow]$ can be understood  if we consider
the proton SkM* spectrum in $^{24}$Mg. Fig.~\ref{24Mg_sps} shows
that, at $\beta $= 0.1-0.3, the Fermi level
(F) $[211\uparrow]$ is close to F-1 state $[220\uparrow]$ and F+1
state $[202\uparrow]$. The resulting large density of states favors the pairing.
With increasing $\beta $, these states are more and more separated, which
decreases and finally destroys the pairing. At 0.5 $< \beta <$ 0.8, the separation
is so large that the pairing vanishes. Following Fig.~\ref{24Mg_sps}, the
transition  energy $E_{ph}$ for configuration pp$[211\uparrow, 202\uparrow]$
(indicated in the figure by arrows) rapidly grows with $\beta $.

The similar analysis shows that, in $^{20}$Ne, the anomalous behaviour of $\cal J$
 is mainly provided by proton and neutron configurations [211$\uparrow$, 220$\uparrow$].

Altogether, we see that, in light deformed Z=N nuclei, a {\it single} 2qp configuration
can determine behavior of ${\cal J}(\beta)$. Moreover, just
this configuration makes the behavior counterintuitive, i.e. with  $d{\cal J}/d\beta < 0$.

\subsection{Search of $d{\cal J}/d\beta < 0$ regime in experiment}

An experimental assessment of the regime $d{\cal J}/d\beta < 0$ requires means to change
the nuclear deformation deliberately. One of the most promising ways
is a nuclear rotation. Soft PES for the nuclei studied here favor
 a change of deformation  already for small angular momenta.

The effect could be searched experimentally by simultaneous
inspection of the intraband energy intervals $\Delta E(I)=E_I-E_{I-2}$
(for determination of ${\cal J}(I)$) and transition probabilities
$B(E2, I \to I-2)\propto Q^2_0 (I) \propto \beta^2(I)$ (for determination
of intrinsic quadrupole moment $Q_0(I)$ and deformation $\beta(I)$) in
the nuclear yrast line $E(I)$.  Then, by plotting ${\cal J}(I)$ and
$\beta(I)$, we should look for $I$-intervals with
\begin{equation}
  \frac{d{\cal J}(I)}{dI} \cdot \frac{d \beta (I)}{dI} <0 \;\; {\rm or} \;\;
  \frac{d{\cal J}(I)}{dI} \cdot \frac{d Q_0^2 (I)}{dI} <0 .
\end{equation}
Both cases correspond to regime $d{\cal J}/d\beta < 0$.

\begin{table} 
\caption{
Experimental \protect\cite{BM2,Hinohara_PRC11,nndc,20Neexp} energies $E_I$, moments of inertia
${\cal J}(I)$, intraband transition probabilities $B_I(E2\downarrow)=B(E2, I \to I-2)$ in the g.s.
 rotational bands, squared intrinsic quadrupole moments $Q^2_0$ and quadrupole deformations
 $\beta$ in $^{24}$Mg and $^{20}$Ne.
 }
\begin{tabular}{|c|c|c|c|c|c|c|}
\hline
Nucleus & $I$ & $E_I$ & ${\cal J}(I)$ & $B_I(E2\downarrow)$
        & $Q^2_0$ & $\beta$ \\
\hline
       & & [MeV] & [$\hbar^2/$MeV] & [${\rm e^2 fm^4}$]  & [$e^2 b^2$] &\\
\hline
           & 2 & 1.368 & 2.19 & 88 & 0.44 & 0.61 \\
 $^{24}$Mg & 4 & 4.122 & 2.54 & 160 & 0.29 & 0.49 \\
           & 6 & 8.113 & 2.78 & 155 & 0.22 & 0.43 \\
\hline
           & 2 & 1.633 & 1.84 & 65.4(32) & 0.33 & 0.71 \\
 $^{20}$Ne & 4 & 4.247 & 2.68 & 70.9(64) & 0.13 & 0.45 \\
           & 6 & 8.777 & 2.43 & 64.5(10) & 0.093 & 0.38\\
\hline
\end{tabular}
\label{exper}
\end{table}

\begin{figure} 
\centering
\vspace{0.5cm}
\includegraphics[width=0.6\textwidth]{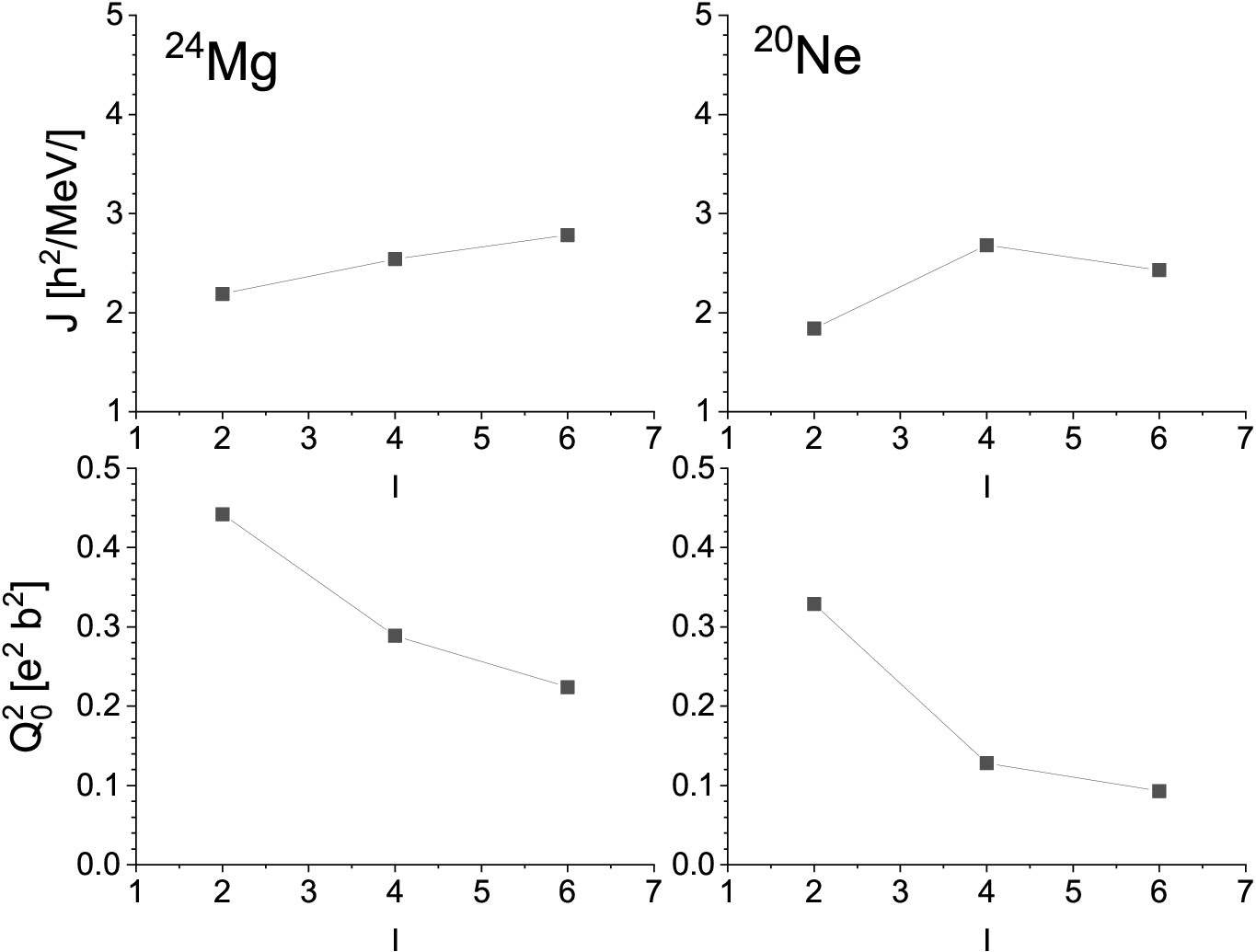}
\caption{Dependence of the experimental moments of inertia $\cal J$ and
squared quadrupole moment $Q^2_0$ on the orbital moment $I$ in
$^{24}$Mg (left panel)and $^{20}$Ne (right panel) \protect\cite{nndc}.}
\label{Jexp}
\end{figure}

In Fig.~\ref{Jexp}, we show ${\cal J}(I)$ and $Q^2_0(I)$ in $^{24}$Mg and $^{20}$Ne,
extracted from the experimental data and listed in
Table~\ref{exper}. The rotational bands in these light nuclei are short. We do not
consider here the terminal rotational states with $I$=8 since available experimental data for
these states are still disputed.

The $\cal J$-values in Fig.~\ref{Jexp} are obtained
from experimental energy intervals in the g.s. rotational band
as \cite{Ring}
\begin{equation}
{\cal J}(I) = \frac{\hbar^2 (2I-1)}{\Delta E_I} .
\label{JIexp}
\end{equation}
For $I$=2, this expression gives the same result as Eq.~(\ref{E_I}).
However, if ${\cal J}(I)$ noticeably changes with $I$,  then Eq.~(\ref{JIexp})
is more relevant for getting ${\cal J}(I)$ at larger spins than direct use
of Eq.~(\ref{E_I}).

$Q^2_0(I)$-values in Fig.~\ref{Jexp} are evaluated using experimental intraband reduced
transition probabilities  $B_I(E2\downarrow)=B(E2, I \to I-2)$:
\begin{equation}
Q_0^2(I)
        =\frac{16 \pi}{5} {\cal R}_I B(E2, I \to I-2) ,
\end{equation}
where
\begin{equation}
{\cal R}_I=\frac{B(E2, I-2 \to I)}{B(E2, I \to I-2)}
   =\left[\frac{C^{I,0}_{I-2,0,2,0}}{C^{I-2,0}_{I,0,2,0}}\right]^2
   =\frac{2I+1}{2I-3}
\end{equation}
and $C^{I,0}_{I-2,0,2,0}$, $C^{I-2,0}_{I,0,2,0}$ are Clebsch-Gordan coefficients.

Fig.~\ref{Jexp} shows that, at $I$=2-6 in $^{24}$Mg and $I$=2-4 in $^{20}$Ne, the values
${\cal J}(I)$ and $Q^2_0(I)$ have opposite trends and so give $d{\cal J}/dI \cdot dQ_0^2/dI <0$.
Note that $^{20}$Ne is deformation-soft and so the effect here can be related to onset of triaxiality
at low spins \cite{Gul22,20Ne_PRC04},
leading to decrease of the axial quadrupole deformation. A decrease  of
$Q^2_0$ with $I$ is indeed seen in Fig.~\ref{Jexp}. This can be accompanied by decrease
of the pairing with $I$.

Appearance of the regime
$d{\cal J}/d\beta < 0$ in the experimental data (Fig.~\ref{Jexp}) is a promising message
in favour of our analysis but not yet its robust proof. Indeed,
our analysis is based on the deformation-constrained calculations
and does not consider a possible dynamical origin (rotation, triaxialy, clustering)
of the effect.

Our calculations omit the Coriolis coupling since it should be weak for the ground-state
bands in $^{24}$Mg and $^{20}$Ne. Indeed, we deal here  with small orbital momenta $I=2-6$. Following
experimental data \cite{nndc} for both nuclei, rotational bands with $K^{\pi}=1^+$, which could be coupled
with g.s. rotational band, are expected only around 10 MeV.

\section{Conclusions}

The evolution of the moment of inertia $\cal J$ with axial
quadrupole deformation $\beta$ in light deformed nuclei
$^{24}$Mg and $^{20}$Ne was investigated in the framework of
macroscopic and microscopic models. For macroscopic treatment,
the rigid-body (RB) and hydrodynamical (HD) models
\cite{BM2,Ring,Sol,Sit} are applied. The microscopic models include
Inglis-Belyaev (IB) \cite{Be59,Be61}, Thouless-Valatin (TV) \cite{TV62}
and linear adiabatic time-dependent Hartree-Fock (ATDHF)
\cite{Ring,ATDHF_GR78,ATDHF_RGG84} approaches, all based on Skyrme
mean-field calculations. The microscopic approaches embrace effects
from shell structure, pairing, and dynamical linear response. The calculations are
performed with three Skyrme parametrizations (SVbas
\cite{SVbas}, SkM* \cite{SkMs} and SLy6 \cite{SLy6}) covering
various isoscalar effective masses $m^*/m$ and different kinds of
pairing.

All the microscopic calculations (IB, TV and ATDHF) predict in
$^{24}$Mg and $^{20}$Ne a maximum in ${\cal J}(\beta)$-dependence
and corresponding counterintuitive
regime $d{\cal J}/d\beta<0$. This  effect is explained by the impact of a {\it single} $1ph$
configuration dominating in $\cal J$
([211$\uparrow$, 202$\uparrow]$ in $^{24}$Mg, and [220$\uparrow$,
  211$\uparrow$] in $^{20}$Ne).
At the deformation range where the effect takes place, the pairing is absent and
the relevant $1ph$
configuration has a particular behavior: its energy grows with
$\beta$ while the cranking single-particle matrix element $\langle q |I_x|q'\rangle$
remains almost constant. This is a fully mean-field effect though
its manifestation becomes possible due to a collapse of the pairing.
Experimental data \cite{Hinohara_PRC11,nndc,20Neexp} for ground-state bands
in $^{24}$Mg and $^{20}$Ne also demonstrate the behavior $d{\cal J}/d\beta<0$  at low spins.
This is a promising message in favor of our analysis.

Our deformation-constrained calculations do not directly
include physical mechanisms (triaxiality, rotation, ...) which could lead to $d{\cal J}/d\beta<0$
regime but focus to a possible simple explanation of this effect in terms of single-particle spectra.
In this connection, note that ${\cal J}(\beta)$-maximum and related $d{\cal J}/d\beta<0$ behavior
was already found in exploration~\cite{Hinohara_PRC11} for $^{24}$Mg,
taking into account the triaxiality and rotation
impacts. Here we provide a possible mean-field  interpretation of such results and show
that a macroscopic behaviour ${\cal J}(\beta)\propto \beta^2$ can be strongly
violated in light deformed nuclei.

\begin{figure} 
\centering
\vspace{0.5cm}
\includegraphics[width=0.6\textwidth]{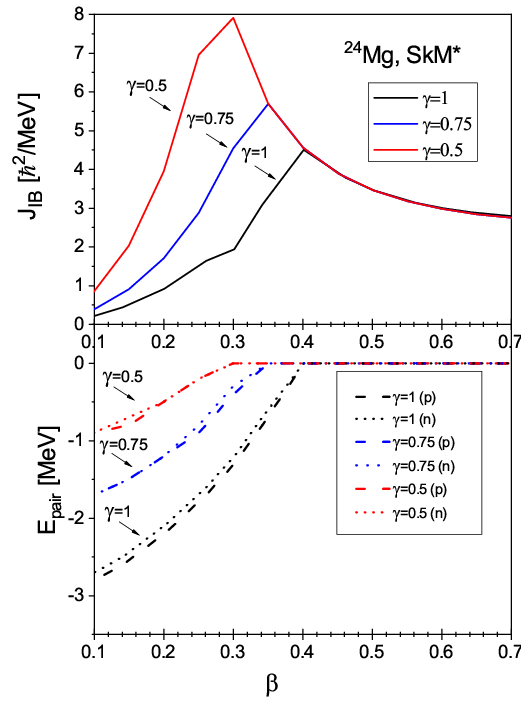}
\caption{ Moments of inertia ${\cal J}_{\rm IB}$ (upper panel) and  proton and neutron pairing energies
(bottom panel) in  $^{24}$Mg, calculated with the force
SkM* at attenuation pairing factors $\gamma$=1, 0.75 and 0.5.}
\label{fig8:weak_pairing}
\end{figure}

\section*{ACKNOWLEDGEMENTS}
\label{s5}

J. K. appreciates the support by a grant of the Czech Science Agency, Project
No. 19-14048S. A. R. acknowledges support by the Slovak Research and Development
Agency under Contract No. APVV-20-0532 and by the Slovak grant agency VEGA
(Contract No. 2/0175/24).

\appendix

\section{Case of a weak pairing}

Though the regime $d{\cal J}/d\beta<0$ is basically the mean field effect, it is actually realized
after the collapse of pairing. In this connection, it is worth to check how the evolution
${\cal J} (\beta)$ would look at the suppressed pairing. We cannot fully turn off the pairing
since
this will lead to unphysically
large values of  ${\cal J}$. Indeed, at weak deformations, the p-h energies entering the
denominator of Inglis models are very small, which, without pairing,
would result in unphysical enhancement of ${\cal J}$. However, we can avoid this problem and see the main
trends if we only partly weaken the pairing, e.g. using the pairing strength constants with attenuation
factors $\gamma$=0.75 and 0.5 (with $\gamma$=1 corresponding to the full pairing).

In Fig.~\ref{fig8:weak_pairing}, the SkM* moments of inertia ${\cal J}_{\rm IB}(\beta)$ in $^{24}$Mg are shown
for $\gamma$=1, 0.75 and 0.5. Besides, the corresponding proton and
neutron pairing energies are exhibited. We see a one-to-one correspondence between the pairing collapse
and onset of the regime $d{\cal J}/d\beta<0$. The weaker the pairing, the smaller the onset deformation $\beta$.
So just the pairing collapse permits the realization of the mean field regime $d{\cal J}/d\beta<0$.

\begin{table*}
\caption{Characteristics of 2qp configurations $qq'$  in $^{24}$Mg
with the largest contributions to ${\cal J}_{\rm IB}$, calculated
with the force SkM* at various deformations. The table includes:
contributions ${\cal J}_{qq'}$ to ${\cal J}_{\rm IB}$, 2qp energies $\epsilon_{qq'}$,
squared matrix elements  $f^2_{qq'}$, Bogoliubov factors $u^2_{qq'}$,
products $(fu)^2_{qq'}$, positions (F-pos) of the s-p levels relative to
the Fermi level (F).}
\begin{tabular}{|c|c|c|c|c|c|c|c|}
\hline
$\beta$ & qq' & ${\cal J}_{qq'}$ & $\epsilon_{qq'}$ & $(fu)^2_{qq'}$ & $f^2_{qq'}$
& $u^2_{qq'}$  & F-pos$_{qq'}$\\ \hline
\hline
0.2 & nn[211$\uparrow$, 202$\uparrow$] & 0.28 & 2.83 & 0.40 & 2.50 & 0.16 & F,F+1 \\ \hline
      & pp[211$\uparrow$, 202$\uparrow$] & 0.26 & 2.95 & 0.38 & 2.53 & 0.15 & F,F+1\\ \hline
      & pp[211$\uparrow$, 211$\downarrow$] & 0.07 & 4.97& 0.17 & 0.28 & 0.61 & F,F+2\\ \hline
      & nn[211$\uparrow$, 211$\downarrow$] & 0.07 & 5.02 & 0.18 & 0.28 & 0.63 & F,F+2\\ \hline
      &  & ${\cal J}_{IB}$ = 0.90 &      &             &             &        &    \\ \hline	\hline
0.4 & nn[211$\uparrow$, 202$\uparrow$] & 1.60     & 2.51 & 2.01	 & 2.43 & 0.83 & F,F+1\\ \hline
      & pp[211$\uparrow$, 202$\uparrow$] & 1.44   & 2.63 & 1.90	 & 2.43 & 0.78 & F,F+1\\ \hline
      & nn[211$\uparrow$, 211$\downarrow$] & 0.19 & 5.20 & 1.00 & 1.07 & 0.94 & F,F+2\\ \hline
      & pp[211$ \uparrow$, 211$\downarrow$]	& 0.19& 5.14 & 0.98 & 1.07 & 0.92 & F,F+2\\ \hline
      &     & ${\cal J}_{IB}$ =	3.75 &      &             &             &        &     \\ \hline	
\hline
0.5 & nn[211$\uparrow$, 202$\uparrow$] & 1.33 & 3.63 & 4.84 & 4.84 & 1 & F,F+1\\ \hline
      & pp[211$\uparrow$, 202$\uparrow$] & 1.33 & 3.64 & 4.84 & 4.84 & 1 & F,F+1\\ \hline
      & pp[211$\uparrow$, 211$\downarrow$] & 0.21 & 5.64& 1.18 & 1.18 & 1 & F,F+2\\ \hline
      & nn[211$\uparrow$, 211$\downarrow$] & 0.21 & 5.80 & 1.19 & 1.19 & 1 & F,F+2\\ \hline
      &      & ${\cal J}_{IB}$ =	3.51 &      &             &             &         &    \\ \hline	
\hline
1.0 & pp[211$\uparrow$,202$\uparrow$]	& 0.64	 & 8.40	& 5.34	& 5.34 & 1	& F,F+1 \\ \hline	
      &nn[211$\uparrow$,202$\uparrow$]	& 0.63	 & 8.44	& 5.31	& 5.31	& 1	& F,F+1 \\ \hline	
      &pp[220$\uparrow$,211$\downarrow$]& 0.23	& 11.30	& 2.62	& 2.62	& 1	& F-1,F+2 \\ \hline	
      &nn[220$\uparrow$,211$\downarrow$]& 0.23	& 11.54	& 2.60	& 2.60	& 1	& F-3,F+2 \\ \hline	
      &pp[101$\uparrow$,330$\uparrow$]	& 0.16	& 4.00	& 0.64	& 0.64	& 1	& F-1, F+6 \\ \hline	
      &pp[211$\uparrow$,440$\uparrow$]	& 0.15	& 11.61	& 1.79	& 1.79	& 1	& F,F+5 \\ \hline	
      &        & ${\cal J}_{IB}$ = 2.95 &      &             &             &        &     \\ \hline			
\end{tabular}
\label{tabMg}
\end{table*}

\section{Properties of 2qp states}

 Table~\ref{tabMg} displays characteristics of 2qp states with the largest
contributions ${\cal J}_{qq'}$  to ${\cal J}_{\rm IB}$ in $^{24}$Mg.
The calculations are performed  with the force SkM* for deformations
$\beta$=0.2, 0.4, 0.5 and 1.0.
It is seen that, for all these deformations, the dominant contribution to ${\cal J}_{\rm IB}$
is provided by proton (pp) and neutron (nn) pairs [211$\uparrow$, 202$\uparrow$].
The domination is explained by three favorable factors: large s-p
matrix element $f_{qq'}$, modest pairing suppression $u_{qq'}$ and
rather low excitation energy $\epsilon_{qq'}$. \cal Just these 2qp pairs are
used for calculation of ${\cal J}^{(2)}_{\rm IB}$.

The large matrix element for $[211\uparrow, 202\uparrow]$
is explained by holding the asymptotic selection rules
$\Delta N=0,\pm2, \Delta n_z=\pm 1, \Delta \Lambda  = 1$ for
$lm=21$-transitions \cite{Sol}.

As seen from the table, the pairing vanishes at $\beta >$ 0.5. Then
we get $u_{qq'}$=1, $\epsilon_{qq'} = |e_p-e_h|$,
and ${\cal J}_{\rm IB} \to {\cal J}_{\rm Ing}$.

\bigskip

\end{document}